\documentclass{aa}  

\usepackage[utf8]{inputenc}
\usepackage{amsmath}
\usepackage{geometry}
\usepackage{verbatim}
\usepackage{indentfirst}
\usepackage{graphicx}
\usepackage{listings}
\usepackage{pythonhighlight}
\usepackage{setspace}
\usepackage{subcaption}
\usepackage{txfonts}
\usepackage{natbib,twoopt}
\usepackage[breaklinks=true]{hyperref} %% to avoid \citeads line fills
\bibpunct{(}{)}{;}{a}{}{,}             %% natbib format for A&A and ApJ
\makeatletter
  \newcommandtwoopt{\citeads}[3][][]{\href{http://adsabs.harvard.edu/abs/#3}%
    {\def\hyper@linkstart##1##2{}%
     \let\hyper@linkend\@empty\citealp[#1][#2]{#3}}}
  \newcommandtwoopt{\citepads}[3][][]{\href{http://adsabs.harvard.edu/abs/#3}%
    {\def\hyper@linkstart##1##2{}%
     \let\hyper@linkend\@empty\citep[#1][#2]{#3}}}
  \newcommandtwoopt{\citetads}[3][][]{\href{http://adsabs.harvard.edu/abs/#3}%
    {\def\hyper@linkstart##1##2{}%
     \let\hyper@linkend\@empty\citet[#1][#2]{#3}}}
  \newcommandtwoopt{\citeyearads}[3][][]%
    {\href{http://adsabs.harvard.edu/abs/#3}
    {\def\hyper@linkstart##1##2{}%
     \let\hyper@linkend\@empty\citeyear[#1][#2]{#3}}}
\makeatother

\newcommand{\Msun}{M$_{\sun}$\xspace}

\begin{document}

\titlerunning{Radio emission from QS Vir}
\authorrunning{Ridder et al.}

\title{Spectral features and variable circular polarisation in the radio emission from the pre-cataclysmic variable QS Vir}

\author{M.~E. Ridder \inst{1}
      \and
      A.~K. Hughes \inst{1, 2}
      \and
      C.~O. Heinke \inst{1}
      \and
      G.~R. Sivakoff \inst{1}
      \and 
      R.~D. Sydora \inst{1}
}

\institute{
        Physics Dept., CCIS 4-183, University of Alberta,
         %116 St \& 85 Ave, 
        Edmonton, AB, T6G 2E1, Canada \\
        \email{mridder@ualberta.ca}
\and
        Astrophysics, Dept. of Physics, University of Oxford,
        Keble Rd,
        Oxford OX1 3RH, United Kingdom \\
}

\date{Received XXX; accepted YYY}

\abstract{QS Vir is a low-accretion rate cataclysmic variable (CV), or pre-CV, as the M dwarf companion is just filling its Roche lobe. We recently identified radio emission from QS Vir in the Very Large Array Sky Survey, at a flux of $\sim$1 mJy. 
The origin of radio emission from CVs is not fully understood, with evidence for synchrotron emission from jets and other coherent plasma emission processes, such as electron cyclotron maser emission (ECME) or plasma radiation.}
{Our aim is to constrain the radio emission mechanism for QS Vir, through spectroscopic, polarisation, and time variability measurements, all while checking for correlated X-ray variations.}
{We took 3 epochs of new observations with the VLA in S, C, and X bands, with full Stokes polarisation information, complemented by near-simultaneous \textit{Swift}/XRT X-ray data.
Radio spectra are extracted to search for emission features characteristic of coherent plasma emission processes (e.g. high circular polarisation and narrow-band emission). We fit the X-ray spectra with absorbed power-laws, finding no strong X-ray variability.}
{QS Vir showed a nearly flat radio spectrum, with fluxes of 0.4-0.6 mJy in all bands. \textit{Swift}/XRT showed $L_X\sim5\times10^{29}$ erg/s in all observations. We identified strong, variable circular polarisation, ranging from 33$\pm$3\% in S band in the last observation, to $<$11\% in the middle observation in all bands. Linear polarisation was not detected, with upper limits of at most 15\%. Intriguingly, the S-band spectra show circularly polarised spectral bumps (width $\sim$0.5 GHz) that rise and decay within $\lesssim$ 5 minutes.}
{We suggest that the radio emission from QS Vir consists of two components: a relatively constant, low-polarisation flat-spectrum component and a band-limited, rapidly variable, and strongly circularly polarised component. This latter coherent component may be associated with ECME or plasma radiation.}

\maketitle

%-------------------------------------------------
\section{Introduction}
Cataclysmic variables (CVs) are interacting binary stars in which a white dwarf (WD) accretes from an M or K type donor that has filled its Roche lobe. The exact geometry of accretion is determined by the WD's magnetic field; this can be via an accretion disk when the WD is weakly magnetic, or onto the poles if the WD is highly magnetic. When the magnetic field is moderately magnetic, there may be a truncated disk. 

A number of CVs have been detected at radio frequencies, across various classes \citep{Benz89,Kording08,coppejans2015,barrett2020}. While radio emission in some appears to be produced by jets \citep{Kording08,Coppejans20}, it is not yet clear for the majority of CVs what the nature of the emission is, or whether the radio emission originates from the accretion process or from the companion star \citep{barrett2020}. Thermal bremsstrahlung emission has been ruled out as a possible radio emission mechanism for other CVs \citep[e.g. ][]{chanmugam1982, kording2008}.
Synchrotron emission from a jet launched by the WD  has been suggested for SS Cyg and other systems \citep{russell2016, mooley2017, fender2019, Coppejans20}. Gyrosynchroton emission (from moderately relativistic electrons) has been proposed for cases such as AM Her \citep{benz1983}.
Coherent plasma emission processes, such as electron cyclotron maser emission (ECME) or plasma radiation, have been argued for numerous CVs \citep[e.g. ][]{barrett2020}. These coherent emission processes can be distinguished from other emission mechanisms by their large circular polarisation fractions, often approaching 100\% \citep{dulk1985}.

\begin{table*}
    \centering
    \caption{The 1--10 keV \textit{Swift} X-ray flux of QS Vir over three epochs. The first three rows show the flux calculated for each epoch, when the photon index was linked for all observations. The last row was calculated assuming that both the flux and photon index were constant for all observations. Errors are 90\% confidence limits. The luminosity was calculated using the Gaia EDR3 distance of \cite{bailerjones2021}, 50.10 $^{+0.07}_{-0.06}$ pc.}
    \label{tab: swift flux}    
    \begin{tabular}{c c c c c c}
    \hline \hline
         Target ID & Date & Flux & Luminosity & Photon Index & 0--10 keV counts \\
         & & $(10^{-12}\text{ erg }\text{s}^{-1}\text{ cm}^{-2})$ & $10^{29}$ erg s$^{-1}$ &  \\
        \hline
         15432 & 2022 Dec 11 & 2.0$^{+1.1}_{-0.8}$ & 6$^{+3}_{-2}$ & 1.6$\pm$0.4 & 20 \\
         15432 & 2022 Dec 31 & 1.6$^{+1.1}_{-0.7}$ & 4$^{+3}_{-2}$ & 1.6$\pm$0.4 & 27 \\
         15477 & 2023 Jan 19 & 1.9$^{+1.8}_{-1.0}$ & 6$^{+5}_{-3}$ & 1.6$\pm$0.4 & 11 \\
         Combined & - & 1.8$^{+0.9}_{-0.6}$ & 5.4$^{+2.7}_{-1.8}$ & 1.5$^{+0.3}_{-0.4}$ & 58 \\
         \hline
    \end{tabular}
\end{table*}

\begin{table*} 
    \centering
    \caption{Stokes I flux (total intensity) of QS Vir over the three epochs. Error values are the RMS of the background. The spectral index (defined as $\alpha$ if $S_\nu \propto \nu^{\alpha}$) was calculated from the S and X band fluxes assuming the central frequencies were 3 GHz and 10 GHz, respectively.}
    \label{tab: qs vir i flux}    
    \begin{tabular}{c c c c c}
    \hline \hline
         Observation date & S band flux & C band flux & X band flux & Spectral index \\
          & ($\mu$Jy) & ($\mu$Jy) & ($\mu$Jy) & \\
        \hline
         2022 Nov 28 & 620$\pm$20 & 532$\pm$15 & 591$\pm$12 & 0.04$\pm$0.03 \\ 
         2022 Dec 22 & 423$\pm$19 & 393$\pm$15 & 425$\pm$11 & 0.00$\pm$0.04 \\ 
         2023 Jan 9  & 645$\pm$19 & 495$\pm$17 & 486$\pm$16 & 0.24$\pm$0.04 \\ 
         \hline
    \end{tabular}
\end{table*}

\begin{table*}
    \centering
    \caption{Stokes V flux (circular polarisation) of QS Vir and corresponding polarisation fractions. Error values are the RMS of the background. The polarisation direction, left circularly polarised (LCP) or right circularly polarised (RCP), is included for flux fits above 3 $\times$ RMS. Fits below this limit were replaced with 3 $\times$ RMS upper limits.}
    \label{tab: qs vir v flux}    
    \begin{tabular}{c c c c c c c}
    \hline \hline
         Observation date & S band flux & S band pol. & C band flux & C band pol. & X band flux & X band pol. \\
          & ($\mu$Jy) & (\%) & ($\mu$Jy) & (\%) & ($\mu$Jy) & (\%) \\
        \hline
         2022 Nov 28 & $-$89$\pm$17 & 14$\pm$3 LCP & 60$\pm$15 & 11$\pm$3 RCP & 56$\pm$11 & 9.5$\pm$1.9 RCP \\ 
         2022 Dec 22 & $<$45 & $<$11 & $<$40 & $<$10 & $<$31 & $<$7 \\
         2023 Jan 9  & $-$211$\pm$19 & 33$\pm$3 LCP & $<$48 & $<$10 & 52$\pm$15 & 11$\pm$3 RCP \\ 
         \hline
    \end{tabular}
\end{table*}

\begin{table*}
    \centering
    \caption{The 3 $\times$ RMS upper limits on total linear polarisation ($\sqrt{Q^2 + U^2}$) of QS Vir.}
    \label{tab: qs vir lin pol}    
    \begin{tabular}{c c c c}
    \hline \hline
         Observation date & S band flux & C band flux & X band flux \\
          & ($\mu$Jy) & ($\mu$Jy) & ($\mu$Jy) \\
        \hline
         2022 Nov 28 & $<$79 & $<$61 & $<$48 \\ 
         2022 Dec 22 & $<$63 & $<$57 & $<$45 \\ 
         2023 Jan 9  & $<$76 & $<$63 & $<$64 \\ 
         \hline
    \end{tabular}
\end{table*}

ECME is a mechanism for generation of coherent radiation in plasmas through direct
amplification of escaping electromagnetic radiation at the electron gyrofrequency or harmonics (in either extraordinary (X) or ordinary (O) modes) via gyroresonant wave-particle interaction. It requires an environment that is diffuse and highly magnetised such that the cyclotron frequency, $f_{ce}$, is greater than the plasma frequency, $f_{pe}$ \citep{dulk1985, treumann2006}. It is known to be produced by planetary magnetospheres (e.g. auroral kilometric radiation in the Earth's magnetosphere), and it is a proposed emission mechanism in solar spike bursts \citep{dulk1985} that produces narrow-band emission at integer multiples of the cyclotron frequency, it serves as a tracer for the magnetic field strength in the emitting region. Plasma radiation, another coherent process, is possible at higher densities ($f_{pe} > f_{ce}$) and is emitted at integer multiples of the plasma frequency, which provides an avenue for measuring plasma density of the emitting region. It is produced in Type III solar bursts \citep{dulk1985} and is the suggested origin for some flares on low-mass stars \citep{stepanov2001}. Both emission mechanisms require a source of energetic, nonthermal electrons, which can be supplied by processes such as magnetic reconnection.

Radio flares generated by ECME and/or plasma radiation are common from low-mass stars and chromospherically active star (CAS) systems, such as BY Dra and RS CVn-type binaries, along with gyrosynchrotron radiation  \citep{dulk1985}. M-type stars have been extensively studied in the GHz range and frequently exhibit radio flares that are highly circularly polarised (often $>40$\%) and vary on timescales from minutes to hours \citep[e.g. ][]{villadsenhallinan2019}. Flares from these stars have been suggested to arise from ECME \citep{gudelbenz1989,villadsenhallinan2019}, or from plasma radiation  \citep{stepanov2001, osten2006}. Indeed, incoherent gyrosynchrotron emission has been shown to be insufficient to produce the brightness temperatures associated with radio emission from some M dwarf stars \citep{pritchard2021}. \cite{yiu2024} observed populations of radio bright stars in the 2--4 GHz Very Large Array Sky Survey \citep[VLASS;][]{vlass} and the Low-frequency Array Two-metre Sky Survey (LoTSS) survey catalog \citep{lotss} that encompass both M dwarfs and CAS systems. They conclude that for their sample, there must be another mechanism other than incoherent gyrosynchrotron emission that could cause the M dwarfs to be more radio luminous than the G\"udel-Benz relationship would predict. For one RS CVn CAS system, HR 1099, \citet{osten2004} argue that the high-frequency radio waves are produced by gyrosynchrotron radiation, while the low-frequency radio emission is from plasma radiation.

Our target, QS Vir, is an 3.6-hour eclipsing low-accretion rate CV, sometimes referred to as a pre-CV, in which the dMe donor has evolved such that it is just filling its Roche lobe \citep{matranga2012}.  QS Vir is likely in the evolutionary stage just before it undergoes Roche lobe overflow (based on the low rotational velocity of the WD, \citealt{parsons2011}). The accretion rate of $1.7\times10^{-13}\text{ M}_\odot \text{ yr}^{-1}$ \citep{matranga2012} is likely produced by an enhanced wind from the donor nearly filling its Roche lobe, but there is no evidence for an accretion disk. At a distance of only 50.1 pc \citep{bailerjones2021}, it is one of the nearest accreting WDs to Earth.

QS Vir was first detected in radio wavelengths by VLASS (2--4 GHz) at a level of 970$\pm$160 $\mu$Jy \citep[2019 April 25;][]{ridder2023}. Currently, only the Quick Look Stokes I images are available from VLASS, providing no polarisation information. In this paper, we present pointed VLA observations from 2--12 GHz, with full Stokes polarisation, and near-simultaneous X-ray observations with \textit{Swift}/XRT. In Section \ref{obser} below, we discuss the observations and analysis. In Section \ref{res}, we present the results of our analysis. Section \ref{disc} compares QS Vir to other stellar radio sources and CVs. We conclude in Section \ref{conc}. Unless otherwise stated, error values in this paper are the root-mean-square (RMS) of a background region in the image.

\section{Observations \& Analysis} \label{obser}

\subsection{Observations}
Our observations were conducted as part of a joint proposal between the Jansky VLA and \textit{Swift}/XRT to search for correlated radio and X-ray variability. We obtained three epochs of 1-hour VLA observations (22B-257) in S, C, and X band (2--4 GHz, 4--8 GHz, and 8--12 GHz, respectively) on 2022 November 28, 2022 December 28, and 2023 January 9. To constrain the mass transfer rate at the time of our VLA observations, we also requested near-simultaneous \textit{Swift}/XRT data. When a VLA observation was taken, we triggered corresponding 1000-second \textit{Swift} X-ray observations, which were completed on 2022 December 11, 2022 December 31, and 2023 January 19. In the first two epochs, the VLA was in C configuration, but it was transitioning from to C$\to$B configuration in the final epoch,  elongating the synthesised beam. 

\subsection{Analysis}

\subsubsection{\textit{Swift}/XRT data}
The \textit{Swift}/XRT spectra were extracted with Xselect \footnote{\url{https://heasarc.gsfc.nasa.gov/ftools/xselect/}} using a 15 arcsecond radius source region. The data were fit with XSPEC \citep{xspec} assuming a power-law spectrum ($N_H$ is effectively zero at this $\sim$50 pc distance). In the final epoch, QS Vir fell on a dead column of \textit{Swift}/XRT's detector, resulting in an increased uncertainty for its X-ray flux. We fit the three observations simultaneously, first assuming QS Vir had the same spectral index in all epochs, then assuming both a stable flux and spectral index between all epochs. Because the 90\% confidence intervals of the first three non-simultaneous fits include the flux of the final simultaneous fit, we have no evidence that QS Vir's X-ray flux varied between observations. Fits for the 1--10 keV flux are presented in Table \ref{tab: swift flux}.

\subsubsection{VLA data}

\begin{figure*}
    \centering
    \includegraphics[width=0.65\textwidth]{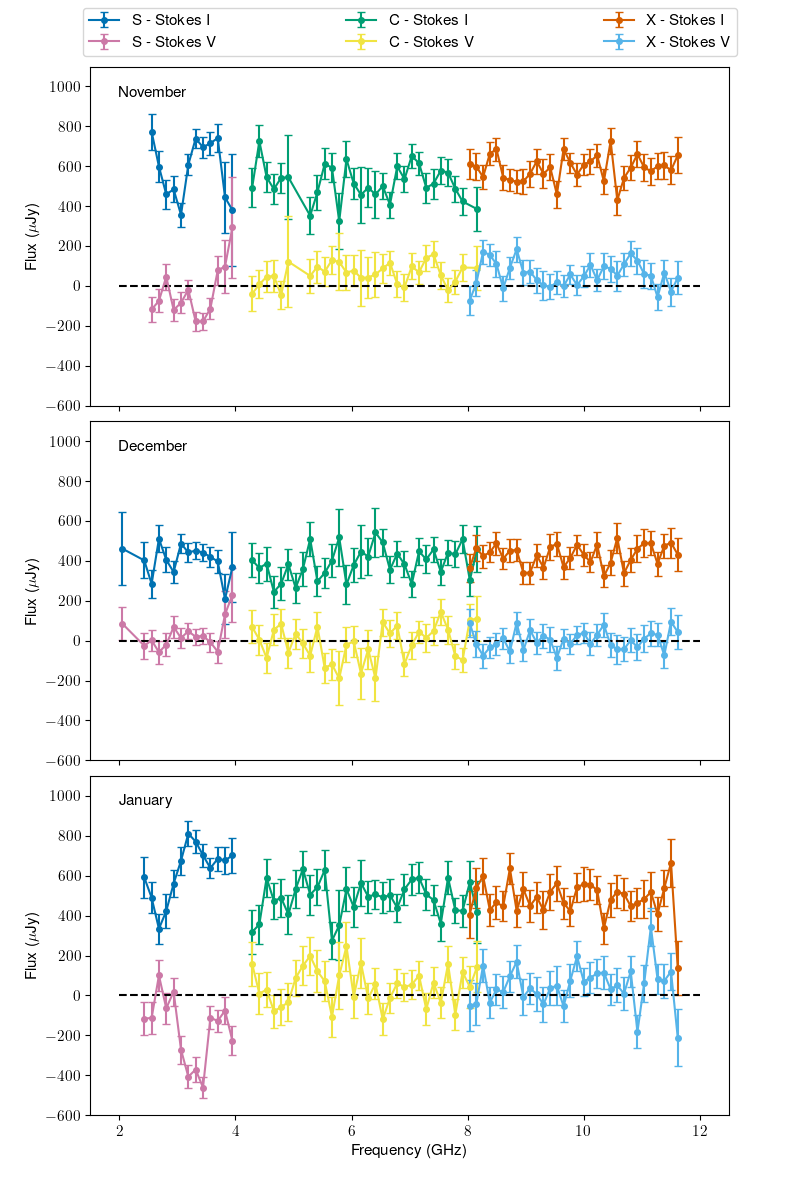}
    \caption{The spectral energy distribution of QS Vir in November, December, and January, which correspond to the top, middle, and bottom panels respectively. Both Stokes I and Stokes V fluxes are presented by spectral window in order to illustrate the frequencies at which we see the most circular polarisation. The fluctuations near the edges of the bands, such as the lower end of S-band and the upper end of X-band, should be interpreted with caution. Large errors and incorrect flux measurements in these ranges are not uncommon.}
    \label{fig: sed}
\end{figure*}

\begin{figure}[ht!]
\centering
\begin{subfigure}{0.45\textwidth}
    \includegraphics[width=\textwidth]{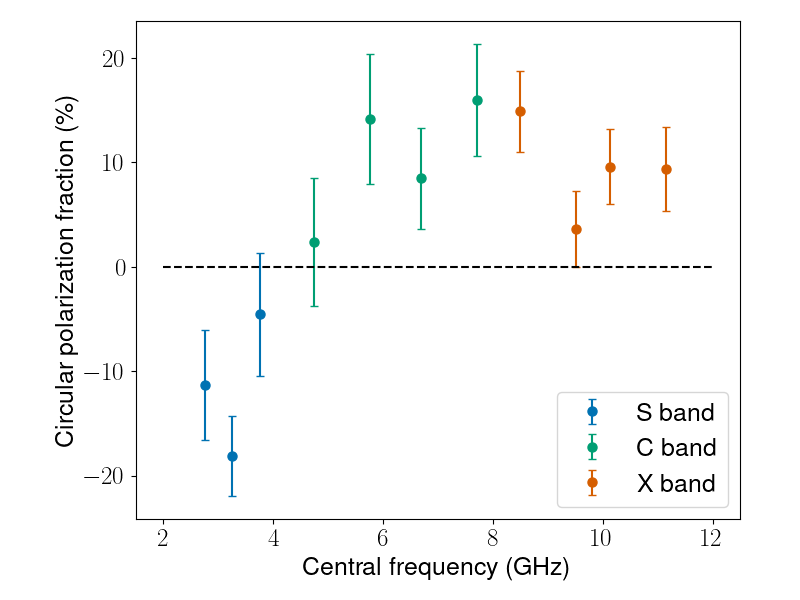}
    %\caption{November}
    \label{fig: nov pol}
\end{subfigure}
\hfill
\begin{subfigure}{0.45\textwidth}
    \includegraphics[width=\textwidth]{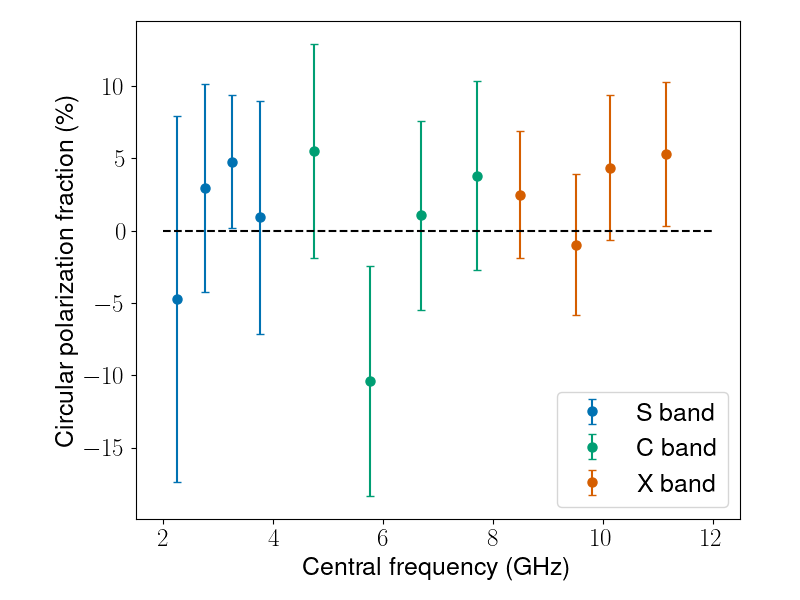}
    %\caption{December}
    \label{fig: dec pol}
\end{subfigure}
\hfill
\begin{subfigure}{0.45\textwidth}
    \includegraphics[width=\textwidth]{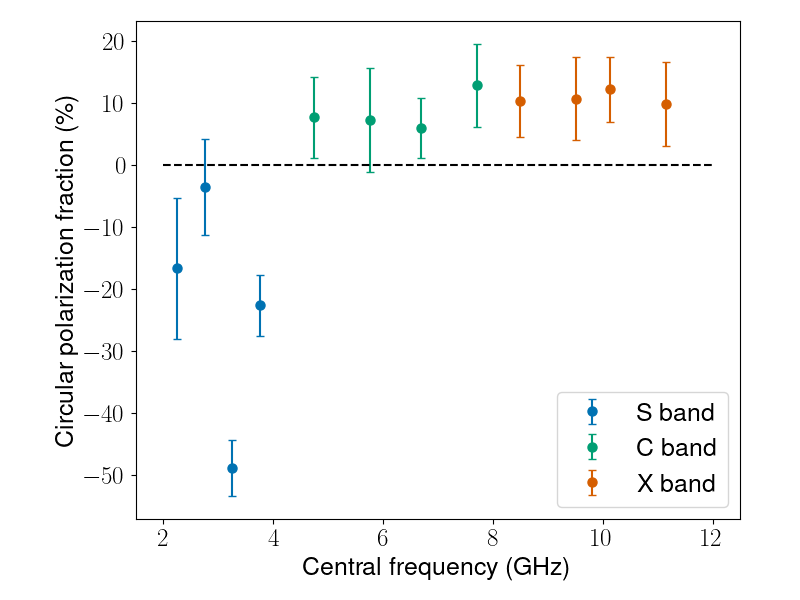}
    %\caption{January}
    \label{fig: jan pol}
\end{subfigure}
\caption{The circular polarisation fraction of QS Vir in all epochs. Each frequency band was separated into 4 sets of spectral windows and imaged separately. The top panel, November, is missing a data point in S band due to the first 4 spectral windows being flagged for RFI. The negative-most point (49$\pm$5\% LCP) in the bottom panel, January, corresponds to the range of the peak frequency in the bottom panel of Figure \ref{fig: sed}(3.2 GHz).}
\label{fig: polfrac}
\end{figure}

Parallel-hand calibration (i.e., Stokes I), automatic flagging, and manual flagging of the VLA data was done with the Common Astronomy Software Applications \citep[CASA; ][]{casa} pipeline version 6.4. We used 3C286 for flux and bandpass calibration, and J1351-1449 as a gain calibrator. Following the pipeline, we applied the cross-hand calibration necessary to measure accurate Stokes Q, U, and V values. Our cross-hand calibration routine was based on the publicly available CASA VLA guides \footnote{\url{https://casaguides.nrao.edu/index.php?title=Karl_G._Jansky_VLA_Tutorials}}; we used our bandpass calibrator to solve for cross-hand phase and J1351-1449 to correct for instrumental leakage. Images in Stokes Q, U, I, and V were constructed with WSClean \citep{wsclean} with masks produced by breizorro \citep{breizorro}.

Despite the sidelobes produced by a bright, steep-spectrum AGN 14' away from QS Vir, self-calibration did not significantly improve the S-band imaging. In the C- and X-bands, the AGN was highly attenuated by the VLA primary beam response.

We fit QS Vir with a 2D Gaussian in CASA 6.6 and report the peak flux (or upper limits) in Tables \ref{tab: qs vir i flux} and \ref{tab: qs vir v flux} for Stokes I and Stokes V, respectively. Upper limits on linear polarisation are presented in Table \ref{tab: qs vir lin pol}.\footnote{The scripts for calibrations, imaging, and fitting have been made available on GitHub: \url{https://github.com/meridder/qs-vir/}.} The spectral energy distributions (SEDs) and polarisation fraction for each epoch are presented in Figures \ref{fig: sed} and \ref{fig: polfrac}. We refer to negative fractions as left circular polarisation (LCP), and positive fractions as right circular polarisation (RCP). We use the convention that RCP is clockwise rotation of the electric field vector as viewed along the direction of propagation and LCP is counterclockwise rotation. We do not detect linear polarisation in any epoch or frequency range of our observations, and provide 3$\times$RMS upper limits in Table \ref{tab: qs vir lin pol}.

Due to the unusually high fraction of circular polarisation in QS Vir, and the change in polarisation handedness between frequency bands (see Tables \ref{tab: qs vir i flux} and \ref{tab: qs vir v flux}), we used the same imaging procedure on the calibrators with the addition of phase-only self-calibration with CubiCal \citep{cubical} to determine if they showed the same polarisation characteristics as QS Vir (details provided in Appendix \ref{appa}). None showed any significant Stokes V flux or Stokes I flux variation. Based on the images of the leakage calibrator, we can estimate the $3\times$RMS upper limit on the instrumental leakage to be $\leq0.4\%$. Sources within the central region of the primary beam, within 90--100\% of the primary beam power, should not be affected by beam squint, which can produce spurious off-axis circular polarisation \citep[see VLA Memo 113; ][]{vla_memo113}. Inside this central beam region (3'), there are no Stokes V sources besides QS Vir, as expected since circularly polarised emission is far more uncommon than linearly polarised emission \citep{yiu2024}. Thus, the calibrators and nearby sources argue for the robustness of our circular polarisation measurements. In Appendix \ref{appb}, we compare our method with alternative gridding algorithms in CASA to verify this result.

\begin{figure}[ht!]
    \centering
    \includegraphics[width=0.45\textwidth]{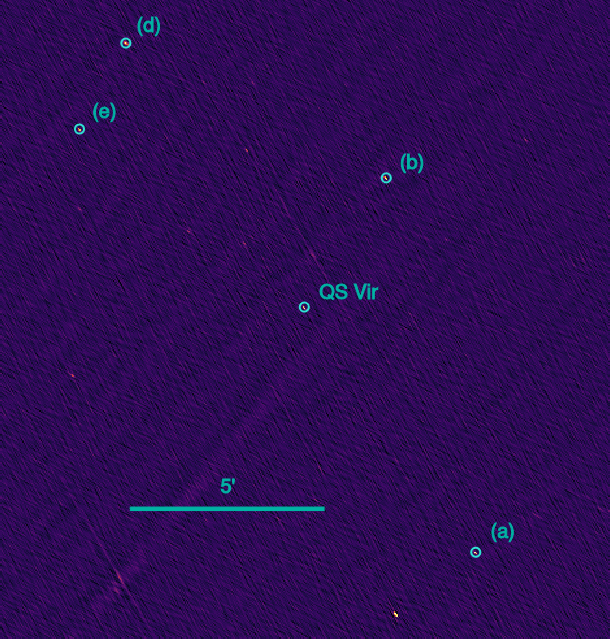}
    \caption{The field of view around QS Vir, including 4 of our check sources, which are labelled to match the panels in Fig.~ \ref{fig: check sources}.}
    \label{fig: field of view}
\end{figure}

\begin{figure}[ht!]
    \centering
    \includegraphics[width=0.45\textwidth]{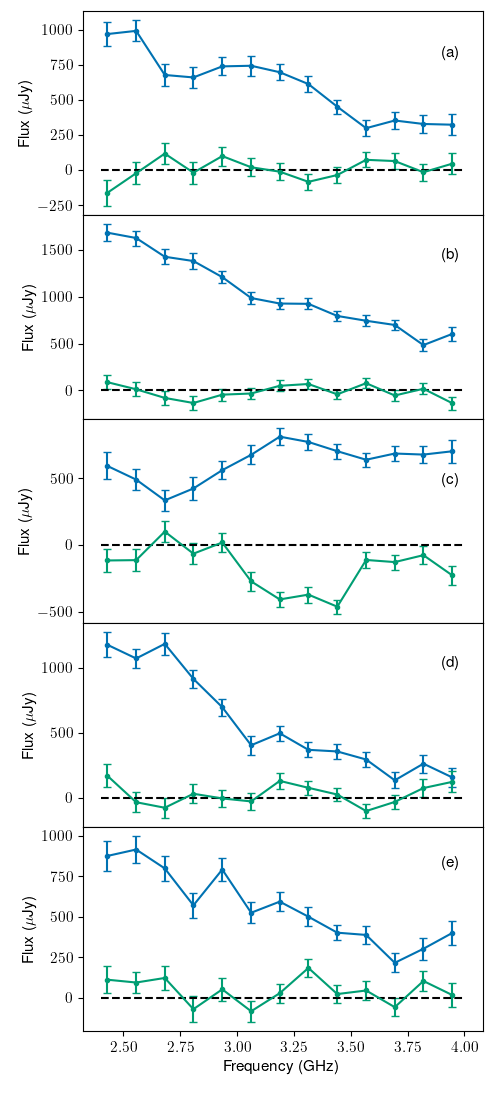}
    \caption{The S-band SEDs of the check sources in Stokes I (blue) and Stokes V (green), compared to QS Vir (panel c) shown in Fig.~\ref{fig: field of view}. Only QS Vir shows significant circular polarisation. 
    }
    \label{fig: check sources}
\end{figure}

\section{Results} \label{res}
The SEDs of QS Vir in Stokes I and Stokes V are shown in Figure \ref{fig: sed}. 
The SEDs are flat, with spectral indices $\alpha$ ($F_{\nu} \propto \nu^{\alpha}$) varying from 0 to 0.24 between epochs (Table \ref{tab: qs vir i flux}). The second epoch is almost 30\% fainter in all bands than the first epoch, while the third epoch recovers in S band, and has a slightly inverted index.

More interesting is the complex spectral shape in S band.  An emission feature is centred around 3.2 GHz in January and 3.5 GHz in November with a width of about 0.5 GHz, peaking at twice the continuum flux. The lowest sub-band in the November data also suggests another emission feature, but we give this data point less credence as it lies at the edge of the band, which suffers from rapid changes in bandpass sensitivity.

Most exciting is the substantial circular polarisation intermittently present in S-band. Figs~\ref{fig: sed} and \ref{fig: polfrac} show strong features in Stokes V around 3--3.5 GHz, matching the emission features in Stokes I. The increase in Stokes V in January is consistent in timing and magnitude with the increase in Stokes I (Fig.~\ref{fig: sed}). Outside the times and frequencies of the flare visible in Stokes I, the S-band continuum appears to have little or no polarisation (see Figs.~\ref{fig: jan flare} and \ref{fig: nov flare}). This indicates that the emission feature is highly circularly polarised, at (conservatively) 80 to 100\%, though the total circular polarisation from 3--3.5 GHz reaches a maximum of 49$\pm$5\% (Fig.~\ref{fig: polfrac}). 

We see varying bumps of circular polarisation at lower levels and specific frequencies in November and January in C and X band (e.g. 5--6 GHz in the January C-band data). Table~\ref{tab: qs vir v flux} shows that the average circular polarisation in C and X bands is above 3 times the RMS in about half the observations.  The tight upper limit on instrumental polarisation leakage argues that these detections are real.  However, there are no corresponding emission features visible in Stokes I in the C and X band data, as is the case in the January S-band data. The November S-band data shows a strong Stokes I emission feature, but a much fainter Stokes V feature, consistent with circular polarisation of the emission feature (subtracting the continuum flux) at up to  50\% ($<20$\% including the continuum; see the top panel in Fig~\ref{fig: polfrac}).

\begin{figure*}
    \centering
    \includegraphics[width=0.8\textwidth]{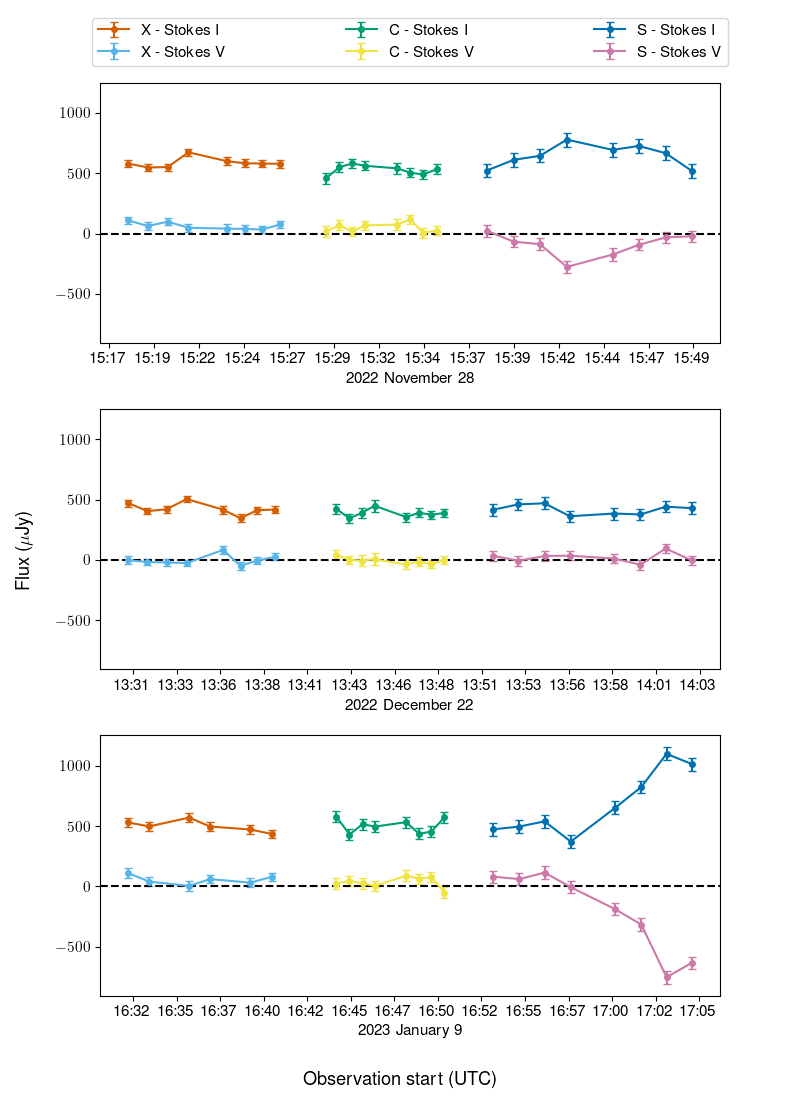}
    \caption{Light curve data for all three epochs and frequency bands in Stokes I and V. We see the most variability in S band in November and January, where increasing Stokes I corresponds to increasingly negative Stokes V.}
    \label{fig: light curve}
\end{figure*}

To verify that these features are real, we measured the fluxes and polarisation of 10 check sources in the field of view, out to $\sim$15'. The positions of 4 nearby sources are shown in Figure \ref{fig: field of view}, and their S band SEDs are shown in Figure \ref{fig: check sources}. None of the check sources show a significant detection in Stokes V. Based on the appearance of these four check sources, there does not seem to be any induced Stokes V in the near field that would contribute to our results with QS Vir. Some off-axis check sources suffer from a higher fraction ($\leq$20\%) of circular polarisation due to beam squint.

\begin{figure}
\centering
\begin{subfigure}{0.45\textwidth}
    \includegraphics[width=\textwidth]{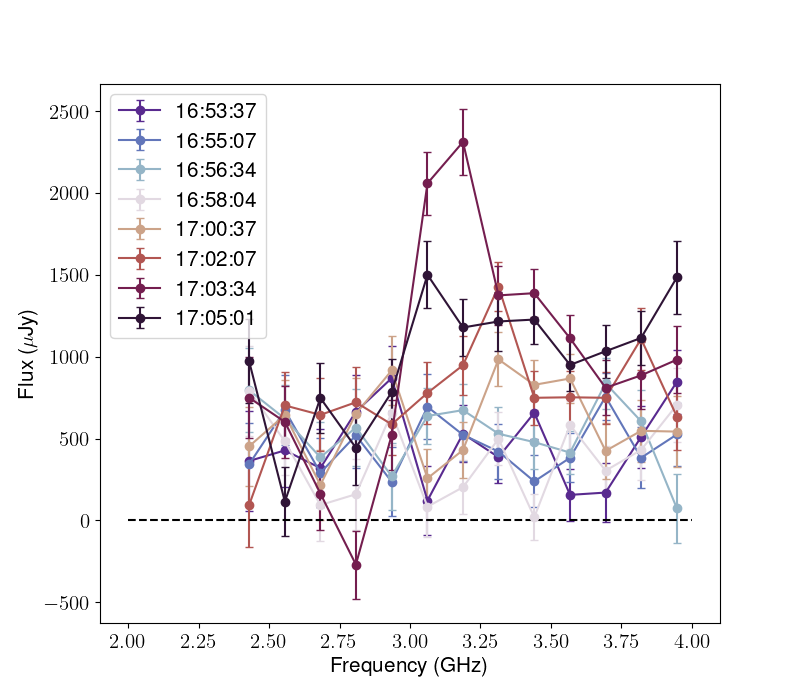}
    \caption{Stokes I}
    \label{fig: jan flare I}
\end{subfigure}
\hfill
\begin{subfigure}{0.45\textwidth}
    \includegraphics[width=\textwidth]{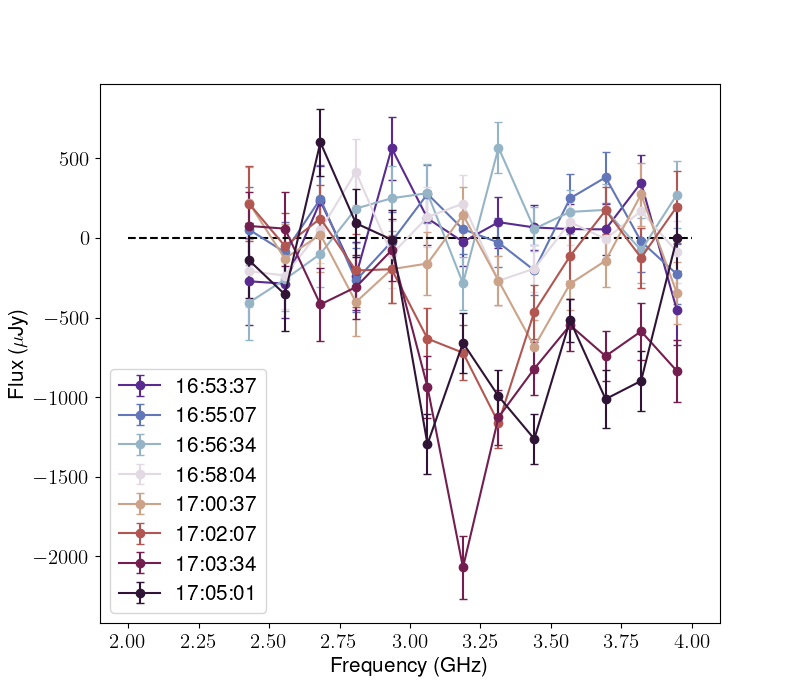}
    \caption{Stokes V}
    \label{fig: jan flare V}
\end{subfigure}
\caption{The S band SED of QS Vir on 9 Jan 2023 in Stokes I ({\it a}) and V ({\it b}), binned in time as in Figure \ref{fig: light curve} with 16 frequency bins. All times are UTC. The spectral feature centered on 3.2 GHz in Stokes I appears at 17:02:07, reaches a maximum at 17:03:34, and decays from 17:05:01. Stokes V shows a flare at the same frequencies and times, with maximum circular polarisation at 17:03:34 (91$\pm$12\%).}
\label{fig: jan flare}
\end{figure}

\begin{figure}
\centering
\begin{subfigure}{0.45\textwidth}
    \includegraphics[width=\textwidth]{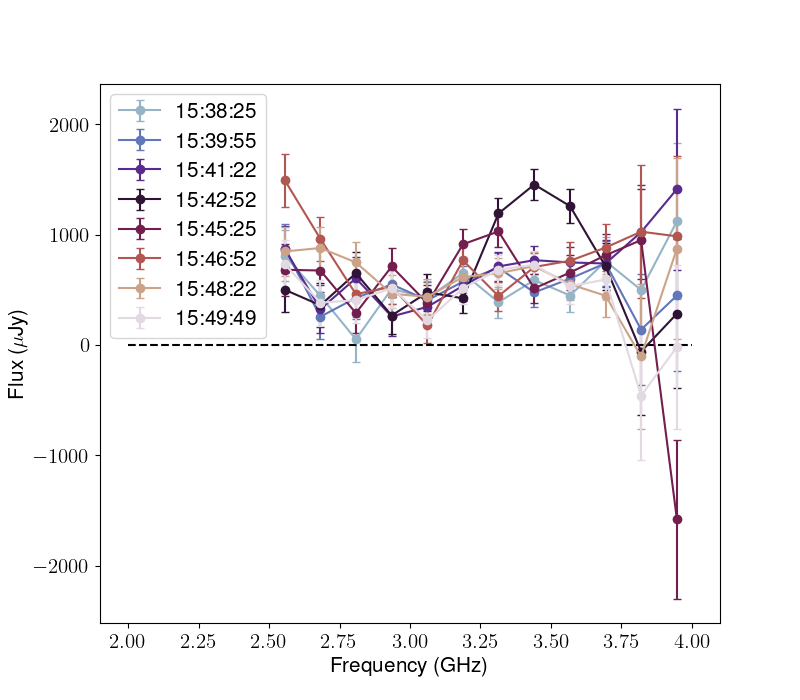}
    \caption{Stokes I}
    \label{fig: nov flare I}
\end{subfigure}
\hfill
\begin{subfigure}{0.45\textwidth}
    \includegraphics[width=\textwidth]{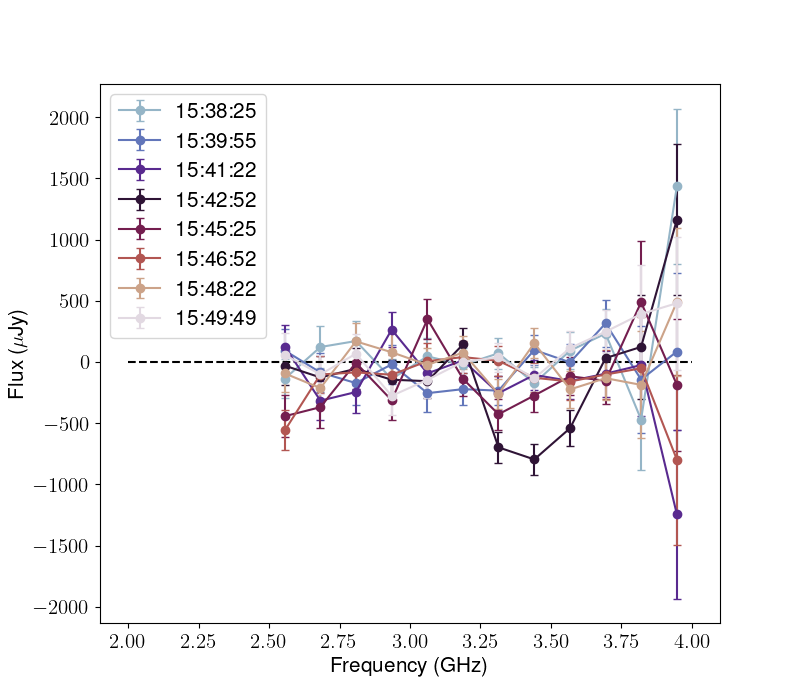}
    \caption{Stokes V}
    \label{fig: nov flare V}
\end{subfigure}
\caption{The S band SED of QS Vir on 28 Nov 2022 in Stokes I ({\it a}) and V ({\it b}), binned in time as in Figure \ref{fig: light curve} with 16 frequency bins. All times are UTC. We do not trust the variation in flux near 4 GHz because it lies near the edge of the frequency band. Similarly to Fig~\ref{fig: jan flare}, a spectral feature centered on 3.5 GHz rises to a maximum at 15:42:52 and decays from 15:45:25. This peak is maximally circularly polarised at 15:42:52 (55$\pm$10\%) and decreases at 15:45:25.}
\label{fig: nov flare}
\end{figure}

We also imaged the data in 4 sub-bands of each frequency band, at resolutions of 0.5 GHz for S band and 1 GHz for C and X band, Figure \ref{fig: polfrac}. The majority of the polarisation signal in January originates from 3--4 GHz. The polarisation fraction between 3 and 3.5 GHz increases the most from 18$\pm$4\% LCP (November) to 49$\pm$5\% LCP (January).

We also split each observation into 8 time bins for each frequency band. The January X band data was split into 6 time bins due to an odd number of scans in this observation. The results in S band of the January epoch (Figure \ref{fig: light curve}) show a flare near the end of the observation that corresponds with increasingly negative Stokes V (i.e., a rise in circular polarisation). The rise time of the peak is 5.5 minutes, over which time the polarisation fraction increased from $<1\%$ to 68\% LCP.  The SEDs for the same time bins in S band are presented in Figure \ref{fig: jan flare}, which show that the increase in flux is due to the same spectral peak as in Figure \ref{fig: sed} rising out of the continuum. A weaker, but qualitatively similar, feature is present in the November S-band data in Fig.~\ref{fig: light curve}, the SED of which is presented in Fig.~\ref{fig: nov flare}. The peak in this epoch, centered on 3.5 GHz, is less obvious, but still corresponds to an increasing fraction of circular polarisation.

Our VLA observations did not include the times of eclipses of the WD, following the ephemeris of \citet{bours2016} and \citet{odonoghue2003}. We compute the accumulated uncertainty in this ephemeris during our observations as only 0.0003 orbital periods. The orbital phases (with eclipses defined as phase 0.0) of the peak flare brightness seen in November and January are 0.8130 and 0.8011, respectively. Our S-band observations of QS Vir cover phases 0.7924--0.8518 in November, 0.5086--0.5680 in December, and 0.7551--0.8146 in January. We find it hard to envision a physical cause for flares to occur at similar orbital phases from November to January, so we tentatively attribute the similarity in phases of these flares to coincidence.

In both Fig.~\ref{fig: jan flare} and Fig.~\ref{fig: nov flare}, we see an indication of a shift in the peak frequency of approximately one 128 MHz bin with each (1.5-minute) time step. The shift is clearer in the Stokes V data for Fig.~\ref{fig: jan flare}. For both epochs, this would correspond to a frequency drift of 1.5 MHz s$^{-1}$, but the poor signal-to-noise in each spectral/timing bin makes this uncertain. This behaviour is reminiscent of frequency drift due to an expanding emission region, which drives the plasma frequency downward. Similar behaviour is observed in Type II solar bursts and is associated with coronal mass ejections \citep{callingham2024}.

In sum, we see flat-spectrum, slowly varying, weakly polarised radio emission from S- to X-bands. On top of this, we see a strong (factor of 4), rapid (minute-scale), circularly polarised (90-100\%) flare, in a narrow (0.5 GHz) band around 3.2 GHz (see Figs.~\ref{fig: jan flare I},\ref{fig: jan flare V}). A second similar, but weaker flare is present in the November S-band data at 3.5 GHz. Smaller levels of circular polarisation in C and X band (Figs.~\ref{fig: sed},\ref{fig: light curve}) may be time-variable and narrow-band, but the signal is not strong enough to make clear statements on this.

\section{Discussion} \label{disc}
\subsection{Comparison of QS Vir to other systems}
To understand the origin of the radio emission from QS Vir, we compare our observations with past observations of other CVs. The CV SS Cyg produces bright flat-spectrum radio emission during outbursts, often attributed to synchrotron emission from a jet, largely due to the radio emission reproducibly tracking the outburst, as seen in the optical (see Introduction; \citealt{russell2016}). QS Vir's accretion rate ($1.7\times10^{-13}$ \Msun/year) is 100 times lower than SS Cyg's inferred accretion rate in quiescence ($\sim10^{-11}$ \Msun/year, \citealt{Wheatley03}), even though QS Vir's radio luminosity ($1.6\times10^6$ mJy$\cdot$kpc$^2$) is slightly higher than the upper limits on SS Cyg's quiescent radio luminosity ($<1.2\times10^6$ mJy$\cdot$kpc$^2$, \citealt{Kording08}). This suggests that the steady radio emission from QS Vir is not produced by the same mechanism generating radio emission during SS Cyg's outbursts. Moreover, synchrotron radiation does not produce highly circularly polarised radio emission, so we exclude synchrotron jet emission as the origin of the highly circularly-polarised narrow-band flares seen in QS Vir.

As summarised in \citet{Kowalski24}, stellar radio flares below 2 GHz tend to be highly circularly polarised, restricted in frequency, and have short durations (minutes). In contrast, higher-frequency stellar radio flares are not significantly polarised, are broader in frequency, and have much longer durations. QS Vir's radio flaring behaviour  resembles, in frequency, duration, and polarisation, what has been seen from several coronally active binaries \citep[e.g.,][]{osten2004,villadsenhallinan2019,Plant24}.

Due to the high fraction of circular polarisation in our observations at 2--4 GHz, we consider ECME or plasma radiation as likely explanations. However, while both mechanisms predict a high fraction of circular polarisation, they do not explain the sign change between S and C band (e.g. the November observations in Table \ref{tab: qs vir v flux}) on their own. Instead, we may be observing two different radiation processes at different frequencies. 

Observations of HR 1099 by \cite{osten2004} show a similar pattern: at low frequencies, the circular polarisation is left-handed, but at higher frequencies, it is right handed. They explain this switch as plasma radiation operating at low frequencies during a flare in HR 1099. The results in \cite{osten2004} show a strong correlation at 20 cm between the overall flux and the circular polarisation fraction, which is similar to our S band data of QS Vir (see Fig.~\ref{fig: jan flare} and \ref{fig: nov flare}). If this were flaring activity akin to that in an RS CVn system, then we would expect an X-ray peak lagging behind the radio by a few hours (the Neupert effect), as seen in HR 1099. Unfortunately, our X-ray observations do not cover this window.

%LCP was less frequently observed at 13 cm (part of S band) than at 20 cm, but their data show fractions from -20\% to -40\% at 13 cm in 1994 August and 1996 September.

Analysis of M-type dwarfs routinely shows similar polarisation fractions in similar frequency ranges to our observations of QS Vir. Observations of AD Leo, a dMe-type star, have repeatedly shown a high degree of circular polarisation \citep[e.g. ][]{gudelbenz1989} up to 100\% RCP. The fine structure of its 2005 flare in L band was compared by \cite{osten2008} to terrestrial auroral kilometric radiation, which is believed to be due to ECME \citep{treumann2006}, concluding that ECME emission is responsible for the flare. 
%They found that given a restricted size on the emission region the maser instability could indeed be possible in the atmosphere of AD Leo. 
\cite{zhang2023} report FAST observations at 1--1.5 GHz of AD Leo that are akin to solar spike bursts with circular polarisation of 35--45\% LCP.  While they do not conclusively identify a physical origin, they suggest ECME.

In addition to AD Leo, \cite{villadsenhallinan2019} report multifrequency (0.3--8.5 GHz) VLA observations of 22 coherent radio bursts of the dMe stars UV Cet, EQ Peg, and YZ CMi. These stars exhibit a mixture of strong left and strong right circular polarisation, depending on the frequency. Although the fraction of circular polarization varies, the handedness as a function of frequency is stable, except in one observation of YZ CMi. Like the \cite{zhang2023} observations of AD Leo, their results show a polarisation fraction of up to 84\% LCP at 0.3--1.6 and 2.8--4 GHz. However, they detect $45\%\pm4\%$ RCP from 1.6--2.2 GHz.

Plasma radiation has been associated with low frequency observations of AD Leo, as for HR 1099. \cite{osten2006} report a flare from AD Leo in the range of 1120--1620 MHz with the Arecibo 305m telescope that shows a circular polarisation fraction of $>90$\% RCP. Higher frequency observations have also been associated with plasma radiation for AD Leo. \cite{stepanov2001} found a polarisation fraction of $\sim100$\% RCP at 4.85 GHz (480 MHz bandwidth), which the authors suggest is plasma radiation despite the relatively high frequency (plasma radiation is often observed in the Sun below 2 GHz).

\begin{figure*}
    \centering
    \includegraphics[width=0.8\textwidth]{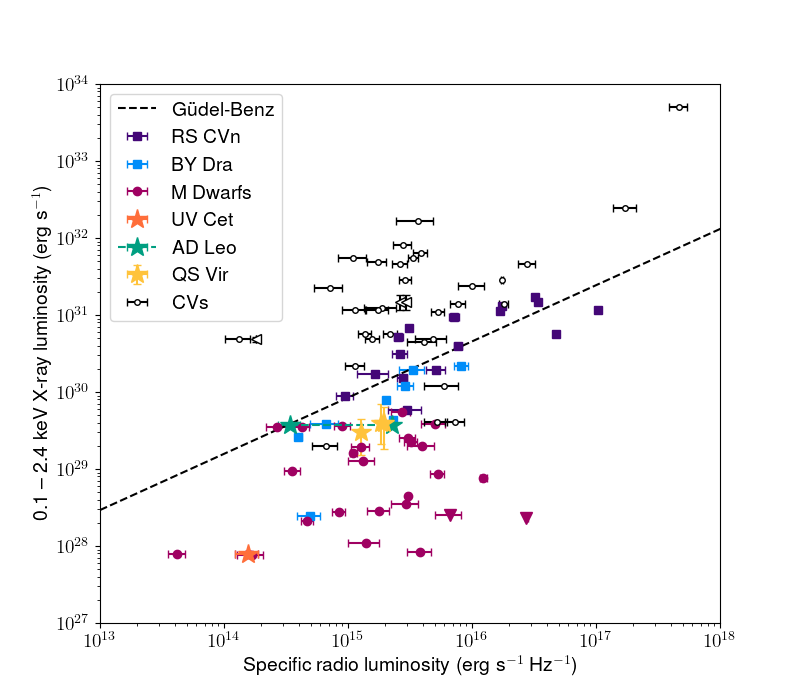}
    \caption{Our S-band QS Vir detections (see Table~\ref{tab: qs vir i flux}) alongside M dwarfs (spectral type M1.5 to M8.5) and CAS systems detected in VLASS from the \cite{yiu2024} dataset as well as AD Leo from \cite{villadsenhallinan2019} on the $L_R$-$L_X$ diagram. We also plot a sample of CVs from \citep{coppejans2015}, \cite{coppejans2016},\cite{barrett2020}, \cite{hewitt2020}, and \cite{ridder2023} that were within 15" of a source in the Second ROSAT All-Sky Survey Catalog \citep{2rxs}. The black dashed line represents the G\"udel-Benz relation \citep{williams2014}. Two M dwarfs from the \cite{yiu2024} sample were excluded from this plot, as their X-ray luminosities were below $10^{26}$ erg s$^{-1}$. UV Cet and AD Leo (discussed above), have also been highlighted. The X-ray luminosities are dominated by systematics, so error bars are not shown, except for QS Vir.}
    \label{fig: lrlx}
\end{figure*}

QS Vir lies between both populations (Figure \ref{fig: lrlx}) among the BY Dra stars and the X-ray bright M dwarfs. However, we know that the majority of X-rays from QS Vir are produced by accretion (due to eclipse observations, \citealt{matranga2012}), implying that QS Vir's stellar flares are radio-luminous compared to the G\"udel-Benz relation (Fig. \ref{fig: lrlx}). The rest of the CV population have higher X-ray luminosities due to their larger accretion rates compared to QS Vir. Results from \citet{callingham2021}, \cite{yiu2024}, and \cite{driessen2024} suggest that a coherent mechanism could be the reason %in their sample of M dwarfs and CAS systems. 
 M dwarfs are radio-luminous compared to the G\"udel-Benz relation. \cite{yiu2024}, which use 2--4 GHz data from VLASS, specifically propose ECME as the mechanism responsible. A coherent mechanism might then also explain QS Vir's radio luminosity. However, objects known to emit coherent radiation do not always deviate from the G\"udel-Benz relation as shown in \cite{vendantham2022}, so this may not be a reliable diagnostic tool for all cases.
%Based on its placement between these populations, we cannot rule out ECME as a possible radio emission mechanism.

As discussed above, chromospheric activity is a source of coherent, circularly polarised radiation. \cite{parsons2011} report the detection of stellar prominences around QS Vir's donor, which %indicates that there could be 
suggests 
chromospheric activity. %on QS Vir's donor star. 
In addition, white light photometry of QS Vir from \cite{odonoghue2003} shows a clear flare profile. Combined with the radio flaring behavior of other dMe stars, the donor star in QS Vir is a good candidate for the source of the radio emission. 
For a magnetic field of 1--4 kG, which should be the case for the dMe star in QS Vir \citep{barrett2020}, the electron cyclotron frequency $f_{ce}$ should range from 3--11 GHz, which corresponds with the frequency range in which we detected circular polarisation. This suggests identifying QS Vir's radio flares with ECME from the donor star. 
Alternatively, the radio emission could be produced near the WD. If we consider a region near the WD in the accretion flow and use the estimate of $\leq$1 MG from \cite{matranga2012}, the electron cyclotron frequency is $\leq$ 3000 GHz.

For the continuum emission, we consider gyrosynchrotron emission from moderately relativistic electrons as a likely origin. %This is the emission mechanism by 
\cite{chanmugam1982} and \cite{benz1983} proposed this emission mechanism for AM Her and SU UMa, respectively. Gyrosynchrotron emission can have some level of circular polarisation %While there can be some level of circular polarisation  depending on the viewing angle 
\citep{dulk1985}, but due to the noise in the SED, we cannot %rule out 
determine 
whether the polarisation fraction measured at 4--12 GHz
is caused by higher order harmonics of the peaks at 3.2 GHz and 3.5 GHz. As mentioned above, these could generate the low level of circular polarisation seen in C-band and X-band. %and can be ruled out for QS Vir due to the size of the emitting region required being larger than the binary.

\subsection{Origin of radio emission in cataclysmic variables}
The identification of weakly polarised, flat-spectrum, slowly varying radio emission, along with highly circularly polarised, narrow-band, rapidly variable radio emission, from QS Vir suggests two separate physical processes. Both emission types are similar to emission seen from coronally active stars and binaries, implying similar mechanisms: a coherent process producing minute-timescale flares and another process producing slowly varying radiation. 

Inspecting the literature on other CVs, we see that it is possible that most or all CVs %not going through outbursts may
show similar behaviour. A few systems have been observed extensively, and have revealed a few bright, short, highly polarised flares; these include EM Cyg \citep{Benz89}, a 7-hour-period dwarf nova CV in outburst; TT Ari \citep{coppejans2015}, a 3.3-hour-period novalike CV, and AM Her \citep{Dulk83}, a 3.1-hour-period polar CV. Including QS Vir , we now have four CVs, spanning a wide range in WD magnetic field strength and mass transfer rate, that produce similarly polarised flares. %This
The lack of correlation between the flare production and WD magnetic field strength or mass transfer rate indicates that the origin of these flares does not depend on the mass transfer rate, nor the magnetic field of the WD, suggesting that these flares may be produced in the chromosphere or corona of the secondary star.
%that the emission region may be elsewhere in these systems.}

A counter-argument is that circular polarisation has been detected in a large number of magnetic CVs from a %shallow (typical 2-minute exposures) 
VLA survey by \citet{barrett2020} with shallow, 2-minute exposures. 
This could suggest that WD magnetic fields relate to the incidence of these flares. 
However, this may be a selection effect; we thus consider the \citet{barrett2020} sample in more detail. 
Of 24 CVs detected with significance $>3\sigma$\footnote{We do not count AE Aqr as a normal CV, but as an unusual, fast propeller system.}, 9-18 show circular polarisation $>66$\%.
Considering the (often sizeable) errors, the fraction of CVs that show significant circular polarisation is 38-75\%. However, this sample is a small fraction of the 120 magnetic CVs observed in this survey, most of which were not detected. 30\% of non-magnetic CVs in the sample of \cite{Coppejans20}, made up of 1-hour exposures, show significant circular polarisation.
We suggest that the higher ratio of circular polarisation seen in the \citet{barrett2020} survey 
%compared to 30\% \citep{Coppejans20} of non-magnetic CVs that show circular polarisation (in generally deeper, 1-hour exposures; \citealt{coppejans2015,coppejans2016}), 
may be a result of the exposure depth, not a preference for systems with magnetic WDs. Short, shallow observations may be typically unable to detect the quiescent weakly polarised emission, but can detect occasional bright polarised flares. Indeed, several of the nearest and most clearly detected CVs in the \cite{barrett2020} sample (AM Her, AR UMa, WX LMi) show little to no circular polarisation. We may see the quiescent weakly polarised emission in these CVs. The fraction of magnetic CVs where circularly polarised emission was seen in 2-minute observations (9-18 of 122, so 7-15\% of the sample) is plausibly reminiscent of the 25\% duty cycle of radio flaring behaviour from five active M dwarfs  \citep{villadsenhallinan2019}, allowing that some of the surveyed CVs are likely too distant for flares to be detected. Deeper VLA exposures of a few magnetic CVs could test this hypothesis.

\section{Conclusions} \label{conc}
Our wide-band radio observations of QS Vir in 3 epochs of data show a continuum from 2--12 GHz that reaches at most moderate circular polarisation (9--11\% RCP). On top of this component, we find two minute-scale flares, which manifest as highly circularly polarised spectral peaks centred on 3.2 GHz and 3.5 GHz, respectively. The peak flux density of the brightest flare at 3.2 GHz (averaged over a 128 MHz bin) was $2300\pm200$ $\mu$Jy and corresponded to a polarisation fraction of 91$\pm$12\% LCP. The polarisation fraction averaged over the entire S-band (2--4 GHz) during the full observation that contained that flare was 33$\pm$3\% LCP. Several other authors have identified circularly polarised emission from CVs \citep{Dulk83, benz1989, coppejans2015, barrett2020}, but our combination of strong signals and broad bandwidths enabled our identification of highly circularly polarised emission that was both rapid temporally and narrow spectrally.

Due to its spectrally flat nature and moderate circular polarisation, we attribute the 2--12 GHz continuum flux to gyrosynchroton emission, as has been suggested for some other CVs by \cite{chanmugam1982} and \cite{benz1983}. However, we cannot clearly establish whether the polarisation observed at our higher frequencies originates from the gyrosynchroton continuum, or from higher order harmonics of the spectral peaks at 3.2 GHz and 3.5 GHz. The short-timescale flaring and strong circular polarisation clearly indicate that a coherent plasma emission process is present. For example, this emission may be generated in the upper atmosphere of QS Vir's dMe donor star, since similar activity and polarisation fractions are observed in other isolated M dwarfs and CAS systems. Because ECME and plasma radiation both produce circular polarisation fractions up to 100\%, we cannot determine the exact emission process. Nevertheless, our results imply that other CVs may also exhibit highly circularly polarised rapid flaring activity in narrow spectral bands. In particular, deeper observations in S-band, both during and outside of CV outbursts, are necessary.

\begin{acknowledgements}
COH is supported by NSERC Discovery Grant RGPIN-2023-04264.
\end{acknowledgements}

\bibliography{references}
\bibliographystyle{aa}

\begin{appendix}
\section{Imaging of the calibrators} \label{appa}
We constructed full-band images of each calibrator to determine whether they showed the same polarisation fractions as QS Vir in each frequency band. To image the calibrators, we used the same method as with QS Vir, but were able to apply phase-only self-calibration using CubiCal. In imaging the flux calibrator, 3C286, we noticed that WSClean struggled to model the synthesised beam, so we enforced a circular beam shape. This should not affect our interpretation of the measured Stokes I and V peak flux, as the background is minimal by comparison.

\section{Comparison to CASA A-projection gridders} \label{appb}
To verify the circular polarisation observed from QS Vir (see Table \ref{tab: qs vir v flux}) is not an imaging artefact introduced by WSClean, we imaged and fit the S-band data from 2023 January 9 using the CASA 6.6 tclean gridders \textit{standard}, \textit{wproject}, \textit{mosaic}, and \textit{awproject}. The final two include an A-projection algorithm to correct off-axis errors introduced by the primary beam pattern (e.g., beam squint resulting from asymmetric primary beam patterns as seen by the right- and left-handed receivers on the VLA). The Stokes I and V fluxes produced by the \textit{standard}, \textit{wproject}, and \textit{mosaic} gridders agree within 1$\sigma$ the results found using WSClean (Table \ref{tab: gridder flux}), but \textit{awproject} was unable to recover the same Stokes V signal despite QS Vir being near the center of the field, where no correction should be needed. The fit of Stokes V, which was fixed at the Stokes I position, was not above the RMS of the background. The bright AGN mentioned in the main text falls very close to the edge of the aperture illumination function applied by \textit{awproject}, which impaired our ability to remove side lobes and enhanced the noise in the image during the cleaning process. As such, we only produced a dirty image with the \textit{awproject} gridder, so that QS Vir was visible in Stokes I.

\begin{table}[hb]
    \centering
    \caption{The S-band Stokes I and V fluxes of the QS Vir data from 2023 January 9 produced by each CASA gridder. The upper limit on the Stokes V flux found by awproject is 3$\times$RMS.}
    \label{tab: gridder flux}    
    \begin{tabular}{c c c}
    \hline \hline
         Gridder & Stokes I & Stokes V \\
          & ($\mu$Jy) & ($\mu$Jy) \\
        \hline
         standard & $640\pm20$ & $-240\pm20$ \\ 
         wproject & $640\pm20$ & $-240\pm20$ \\
         mosaic  & $640\pm60$ & $-240\pm40$ \\
         awproject  & $660\pm30$ & $< 66$ \\
         \hline
    \end{tabular}
\end{table}

\end{appendix}

\end{document}